\documentclass[]{article}
\usepackage[utf8]{inputenc}
\usepackage{amssymb}
\usepackage{amsmath}
\usepackage{amsfonts}
\usepackage{mathrsfs}
\usepackage{anysize}
\usepackage{epsf}
\usepackage{graphicx}
\marginsize{3cm}{3cm}{2cm}{3cm}

\begin{document}
\title{The Application of Weierstrass elliptic functions to Schwarzschild Null Geodesics}
\author{G. W. Gibbons$^{1,2}$ and M. Vyska$^{2}$  
\\
\\   \small{1. D.A.M.T.P., University of  Cambridge, 
Wilberforce Road, Cambridge CB3 0WA, U.K.}
\\   \small{2. Trinity College, Cambridge, Cambridge CB2 1TQ, U.K. }  
\\}
\maketitle

\begin{abstract}
In this paper we focus on analytical calculations involving null geodesics in some spherically symmetric spacetimes. We use Weierstrass elliptic functions to fully describe null geodesics in Schwarzschild spacetime and to derive analytical formulae connecting the values of radial distance at different points along the geodesic. We then study the properties of light triangles in Schwarzschild spacetime and give the expansion of the deflection angle to the second order in both $M/r_0$ and $M/b$ where $M$ is the mass of the black hole, $r_0$ the distance of closest approach of the light ray and $b$ the impact parameter. We also use the Weierstrass function formalism to analyze other more exotic cases such as Reissner-Nordstr\o m null geodesics and Schwarzschild null geodesics in 4 and 6 spatial dimensions. Finally we apply Weierstrass functions to describe the null geodesics in the Ellis wormhole spacetime and give an analytic expansion of the deflection angle in $M/b$.
\end{abstract}

\thispagestyle{empty}
\newpage
\setcounter{page}{1}
\section{Introduction} Geodesics in Schwarzschild spacetime have been
studied for a long time and the importance of a good understanding of
their behavior is clear. In this paper we shall  focus on analytical
calculations involving null geodesics.  While these are interesting in
their own right, calculations  like this are also important for
experiments testing General Relativity to high levels of
accuracy. Examples of two such proposed experiments are  "The Laser
Astrometric Test of Relativity" or LATOR and "Beyond Einstein Advanced
Coherent Optical Network" or BEACON, which are both using paths of
light rays to verify General Relativity
 and are described in detail in \cite{exptest}.
Both are intended to measure second order effects in  light bending 
Elliptic functions have been used to describe the geodesics in
Schwarzschild spacetime before, mainly in \cite{Hagihara} and  more
recently in \cite{Dabrowski,Hackmann}. In \cite{Hagihara,Hackmann}
the  focus is  mainly on the paths of
massive particles and even though they mention the possibility of
using Weierstrass functions in the null case, they don't go into much
detail. The discussion in \cite{Dabrowski} 
is concerned with null geodesics 
around a charged neutron star using the Reissner-Nordstr\o m metric,
a case we also study but with a different emphasis.
In this paper we  begin by  providing   a complete description of 
Schwarzschild null geodesics in terms of Weierstrass functions and
then, and this is our principal innovation, 
using various ``addition formulae'' for 
Weierstrass functions \cite{Whittaker},  we derive some 
analytical formulae connecting values of
radial distance at different points along  the geodesic. The motivation
is to develop, as far as is possible,   optical trigonometry
in the presence of a gravitating object such as a star or a black hole.  
To that end we  use these the additon formulae to 
study  the properties of light triangles in the
Schwarzschild metric and obtain the deflection angle of the scattering
geodesics to second order in both $M/r_0$ and $M/b$ where $M$ is the
mass of the black hole,  $r_0$ the
distance of closest approach of the light ray and $b$ the impact
parameter.\\ In the final section  we  show how the same  methods
to treat  null geodesics in more exotic spacetimes;
charged black hole, the Ellis wormhole \cite{DeySen}  and Schwarzschild black holes in 4
and 6 spatial dimensions. Although not a primary  concern of the present paper,
it is worth remarking that the addition formulae 
for Weierstrass functions that we make use of
are  closely related to the  existence of an abelian group 
multiplication law on 
any  elliptic curve \cite{McKean} and suggest, in view of the
importance of the complex black hole spacetimes  at the quantum level, 
that it might prove fruitful to
explore this aspect of the theory further.

The organization of the paper is as follows. In section 2 of this paper, we provide the full solution for Schwarzschild null geodesics in terms of Weierstrass elliptic functions and apply it to obtain addition formulae connecting three points on the geodesic. We then calculate the deflection angle of the scattering geodesics to second order in both $M/r_0$ and $M/b$ where $M$ is the mass of the black hole,  $r_0$ the distance of closest approach of the light ray and $b$ the impact parameter. The section is concluded with the discussion of the light triangles and Gauss-Bonnet theorem. In section 3, we apply the Weierstrass function formalism to further examples such as Reissner-Nordstr\o m null geodesics and Schwarzschild geodesics in more spatial dimensions. At the end of the section, we give a detailed description of the Ellis wormhole null geodesics. 

\section{Schwarzschild null geodesics}
The equation obeyed by  a  null geodesic 
$r(\phi)$ in the Schwarzschild metric is 
\begin{equation}
\left(\frac{dr}{d\phi}\right)^2 = Pr^4 -r^2 +2Mr \label{eqn}\,, 
\end{equation}
where $P = E^2/L^2 = 1/b^2$. Here $E$ is the energy of the light, $L$ the angular momentum and $b$ the impact parameter. Interestingly the same
equation  arises for a null geodesic $r(\phi)$ in the Schwarzschild-de-Sitter
or Kottler  metric \cite{Islam}\cite{GWW} and many  of our results
remain vaild in that case. Geometrically, one may regard solutions of 
(\ref{eqn}) as unparameterised geodesics of the optical metric
 \begin{equation}
 ds_o^2 = \frac{dr^2}{\bigl(1-\frac{2M}{r} \bigr ) ^2 } + 
\frac{r^2}{1-\frac{2M}{r}}
\bigl( d \theta ^2 + \sin^2 \theta  d \phi ^2 \bigr) \,,  \label{opt}
\end{equation}
with $\theta= \frac{\pi}{2}$.
Introducting the isotropic coordinate  
$\rho=\frac{1}{2} (r-M ) +\frac{1}{2}  \sqrt{r(r-2M)}$, we find that
 \begin{equation}
 ds_o^2 = n^2(\rho) 
 \Bigl\{ d \rho^2 +\rho ^2 \bigl( d \theta ^2 + \sin^2 \theta d \phi ^2 \bigr) \Bigl\} \,,  
\end{equation} where
 \begin{equation}
n(\rho)= \frac{\bigl(1+\frac{M}{2 \rho} \bigr ) ^3}
{\bigl(1- \frac{M}{2 \rho} \bigr)} \,. 
\end{equation} Thus our results also apply to 
light rays moving in  an isotropic
but inhomogeneous  optical medium in flat space with refractive index
$n(\rho)$.   

Another interpretation of  (\ref{eqn}), recently exploited in
\cite{Pretorius}, is provided by substituting $r=\frac{1}{u}$ in  
and differentiating to obtain
\begin{equation}
\frac{d^2u}{d \phi ^2} +u  = \frac{1}{h^2 u^2} F(u)\,, 
\label{newton}\end{equation}
with
 \begin{equation}
F(u)  = 3M h^2 u ^4 \,.
\end{equation}
Now (\ref{newton}) is the equation governing the motion
of a non-relativistic particle of angular momentum per unit mass
$h$ moving under the influence of a central force $F(u)$.  
In our case the effective force $F(u)$ is attractive, and varies inversely
as the fourth power of the distance. A search of the  voluminous
nineteenth century literature on such problems reveals
that it was comparatively well known that although this problem admits
some simple exact  solutions, which  we shall detail below,
the general solution requires elliptic functions.  
  
If we had adopted isotropic coordinates and substituted
$\rho = \frac{1}{u}$  we would have obtained a very different
formula for $F(u)$. In fact in that case we would have 
 
 \begin{equation}
F(u)= 2M u^2 \frac{(1+ \frac{Mu}{2}) ^5(1- \frac{Mu}{4})} {(1-\frac{Mu}{2}) ^3 } \,. 
\end{equation}

As we shall see in detail in  a later section, the null geodsics
of neutral Tangherlini black holes in $D$ spacetime dimensions
correspond, in Schwarzschild coordinates,
to the motion of a non-relativistic particle with a force
$F(u) \propto r^{-D}$. The cases $D=4,5,7$ are the only cases
known to be integrable in terms of elliptic functions.   
In fact the cases $D=4$ and $D=7$ may be related by a conformal mapping
introduced in his context by Bohlin \cite{Bohlin}  and elaborated upon by 
Arnold \cite{Arnold}. The Bohlin-Arnold mapping is  a type of duality,
i.e. it is involutive,  and the case $D=5$ is self-dual.

\subsection{Weierstrass functions solution}
Substituting $y = M/2r -1/12$  in (\ref{eqn})  gives
\begin{equation}
(y')^2 = 4y^3 -\frac{1}{12}y -g_3 \,,
\end{equation}
where 
\begin{equation}
g_3 = \frac{1}{216} -\left(\frac{M}{2}\right)^2P\,. 
\end{equation}  
In the case $g_3\not =\pm 1/216$ the general solution to this equation 
is $y(\phi) = \wp(\phi + C)$ where $\wp(z)$ is Weierstrass elliptic 
function and $C =$ const. Detailed description of these functions 
together with the proof of the above statement can be found 
in \cite{Whittaker} pages 429-444 and page 484. 
For the critical values of $g_3$ the equation for $r$ can be integrated to 
give
\begin{equation}
r(\phi) = M(1+\cos(\phi))
\end{equation}
in the case $g_3 = 1/216$ and
\begin{equation}
\frac{M}{r(\phi)} = \frac{1}{3} - \frac{1}{1\pm\cosh(\phi)} 
\,, \end{equation}
in the case $g_3 = -1/216$. The former, geometrically  a 
{\it cardioid} in $(r, \phi) $ coordinates,  starts at the singularity, 
reaches the horizon from below and then returns back. 
The latter describes two types of trajectories, 
one starting at infinity, the other at the singularity and both 
approaching the photon sphere, never reaching it.\\ 
Now suppose that $g_3\not =\pm 1/216$ and $M^2P<1/27$. 
Then the polynomial $4y^3 - y/12 - g_3$ has 3 real roots 
$e_1>e_2>e_3$ and the half-periods of the corresponding 
Weierstrass function $\wp$ are
\begin{equation}
\omega_1 = \int_{e_1}^{\infty}\frac{dt}{\sqrt{4t^3 -t/12 -g_3}}\,, 
\end{equation}
\begin{equation}
\omega_3 = -i\int_{-\infty}^{e_3}\frac{dt}{\sqrt{g_3 +t/12 - 4t^3}} \,, 
\end{equation}
where $\omega_1\in\mathbb{R}$ and $i\omega_3\in\mathbb{R}$. 
In this case $\wp(z)$ is real on a rectangular grid with vertices 
$0,\omega_1,\omega_3,\omega_1 + \omega_3$ and 
since $y(\phi)$ is real, the only physical solutions to (2) are 
$y(\phi) = \wp(\phi + \phi_0)$ or $y(\phi) = \wp(\phi + \phi_0 + \omega_3)$
 where $\phi_0\in\mathbb{R}$. We have two cases:\\\\
\textbf{(i) Scattering paths}\\
The point $r=\infty$ corresponds to $y = -1/12$ and 
$\wp(z)$ takes value $-1/12$ at $z$ such that Im$(z) = \omega_3$. 
Therefore, choosing line $\phi=0$ to be the axis of symmetry, 
we have $y(\phi) = \wp(\phi+\omega_1+\omega_3)$ and so
\begin{equation}
\frac{M}{r(\phi)} = \frac{1}{6} +2\wp(\phi+\omega_1+\omega_3) \,,
\end{equation}
where of course the function $\wp$ depends on $P$. Here the 
range of $\phi$ is $[-\beta,\beta]$ where 
\begin{equation}
\beta = \omega_1 - \int_{e_3}^{-1/12}\frac{dt}{\sqrt{4t^3 -t/12 -g_3}} \,. 
\end{equation}
In this notation the angle of deflection $\delta\phi$ is 
$\delta\phi = 2\beta -\pi$.\\\\
\textbf{(ii) Trapped paths}\\
These begin and end at the singularity and $r=0$ corresponds to 
$y = \infty$. So, choosing line $\phi=0$ to be the axis of 
symmetry once again gives $y(\phi) = \wp(\phi +\omega_1)$ where $\phi\in[-\omega_1,\omega_1]$. So in this case
\begin{equation}
\frac{M}{r(\phi)} = \frac{1}{6} + 2\wp(\phi + \omega_1) \,.
\end{equation} \\\\
Figure 1  shows the argument of 
$\wp$ in the scattering and trapped cases. 

\begin{figure}[htbp]
  \centering
    \includegraphics[scale=0.6]{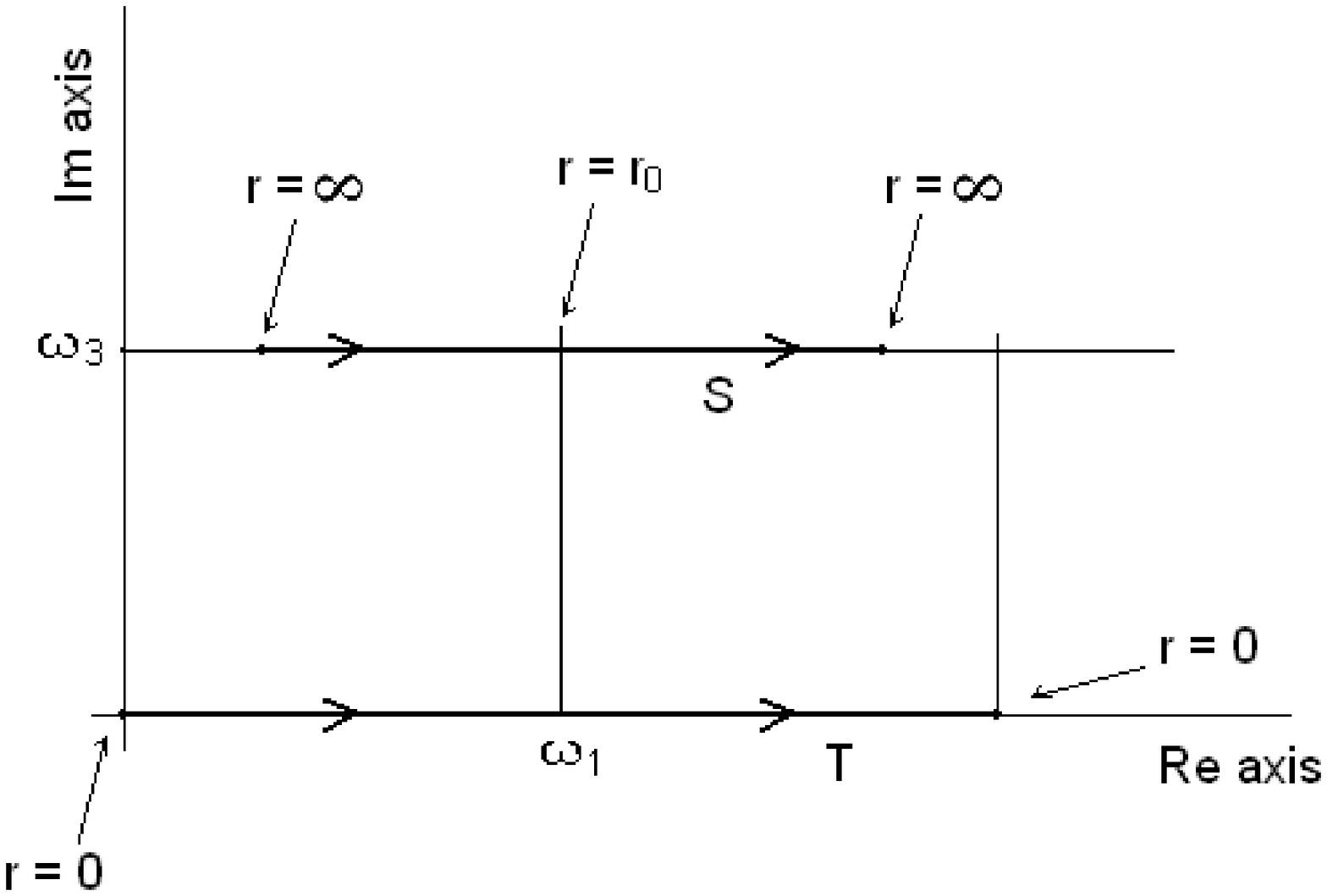}
  \label{Figure}
  \vspace{3mm}
  \centerline{\textbf{Figure 1} \emph{The argument of $\wp$  in the complex plane, S corresponds to scattering trajectories, T to trapped ones}}
\end{figure}

Now suppose that $M^2P>1/27$. Then we have\\\\
\textbf{(iii) Absorbed paths}\\
These go from infinity to $r=0$ or from $r=0$ to infinity and 
therefore the solution is uniquely determined by $P$. 
There is only one real root of the 
r.h.s.  of Weierstrass equation, $e_1<-1/12$ and $\omega_1$ 
defined as before is again a half-period. For each $P$ there is a 
solution of the form $y(\phi) = \wp(\phi +\phi_0)$, $\phi_0\in\mathbb{R}$ 
and so by uniqueness, all physical solutions are of this form.\\ 
We can take $\phi_0=0$ which means defining the line $\phi=0$ by the 
direction in which the path leaves/hits $r = 0$. 
Then the range of $\phi$ is $[-\alpha,\alpha]$ where 
\begin{equation}
\alpha = \int_{-1/12}^{\infty}\frac{dt}{\sqrt{4t^3 -t/12 -g_3}} \,, 
\end{equation}
and the solution is
\begin{equation}
\frac{M}{r(\phi)} = \frac{1}{6} + 2\wp(\phi)\,.
\end{equation}
A diagram of the complex plane  corresponding to absorbed 
trajectories may  be found in the Figure 2.

\begin{figure}[htbp]
  \centering
    \includegraphics[scale=0.61]{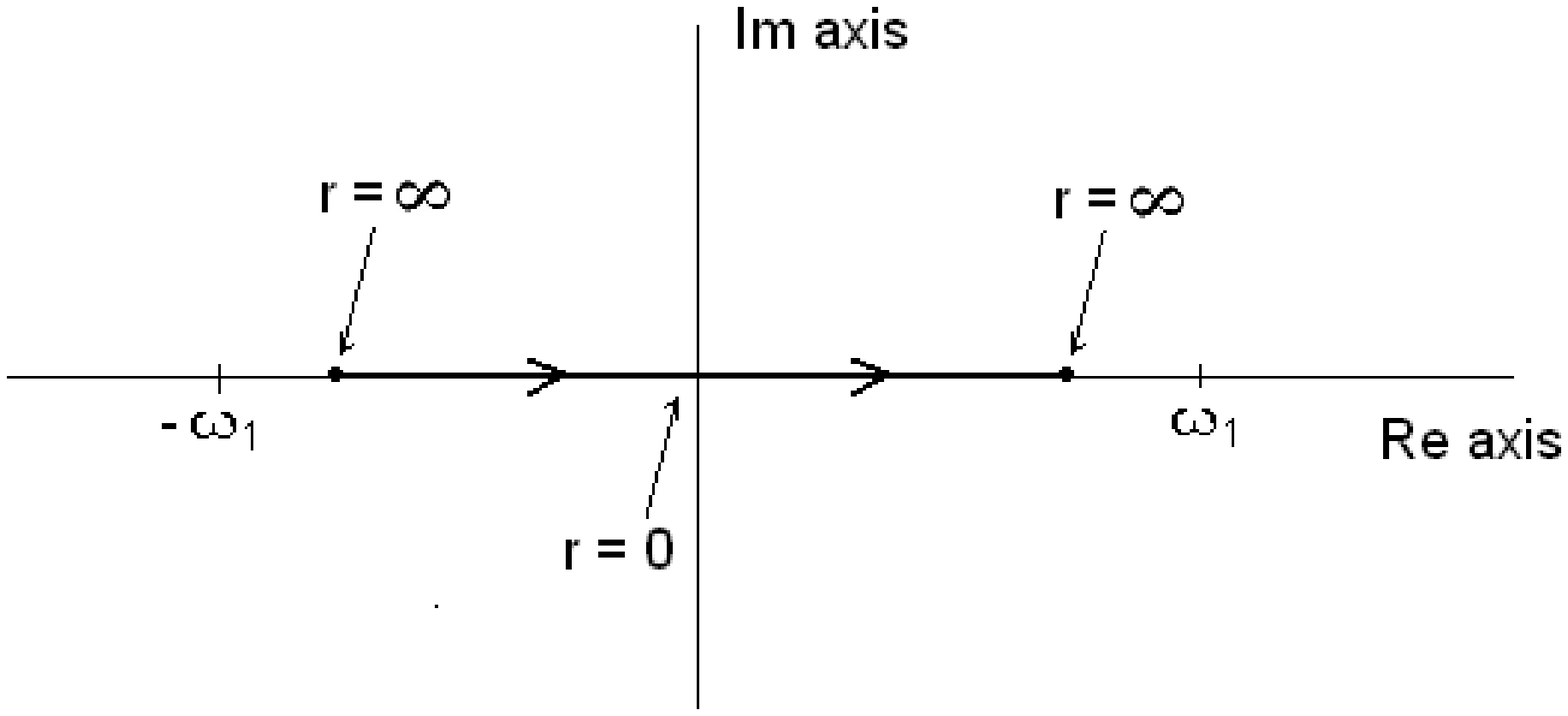}
  \label{Figure}
  \vspace{3mm}
  \centerline{\textbf{Figure 2} \emph{The argument of $\wp$  in the 
complex plane corresponding to absorbed trajectories}}
\end{figure}

\subsection{Addition formulae}
As shown in \cite{Whittaker} page 440, Weierstrass functions satisfy an addition formula of the form 
\begin{equation}
\wp(x+y) = \frac{1}{4}\left[\frac{\wp '(x) - \wp '(y)}{\wp(x)-\wp(y)}\right]^2 -\wp(x) -\wp(y) \equiv F(\wp(x),\wp(y)) \,, 
\end{equation}
where
\begin{equation}
F(x,y) = \frac{1}{4}\left[\frac{\sqrt{4x^3-x/12-g_3} - \sqrt{4y^3-y/12-g_3}}{x-y}\right]^2 - x - y \,.
\end{equation} We can apply this result to null geodesics to obtain an
expression for $r(\phi_1+\phi_2)$ as a function of $r(\phi_1)$ and
$r(\phi_2)$.  Of course, if any such formula is to be useful in some
experimental setup,  we need to be able to easily find the line
$\phi=0$. Also,  because of the additive constant in the argument of
the  Weierstrass function, we cannot apply the addition formula
directly  because the sum of the two arguments will not correspond to the
sum of the  two angles. Fortunately, in the case of the scattering and
trapped orbits,  choosing the line $\phi=0$ to be the axis of symmetry
takes care of both problems. Take the scattering orbit for example.
The axis of symmetry is easy to find, and we can apply the addition formula
for  $\wp$ to 3 points on the orbit $y_1 =y(\phi_1)$, $y_2 =
y(\phi_2)$ and  $y_3=y(\phi_3)$ as 
\begin{equation}
y(\sum_{i=1}^3 \phi_i) = \wp(\sum_{i=1}^3 \phi_i +3\omega_1+3\omega_3) = 
F(F(y_1,y_2),y_3) 
\end{equation}
which works because $2\omega_1$ and $2\omega_3$ are periods of $\wp$.\\
Now, letting $\phi_3 = 0$ gives $y_3 = e_2$ with $e_2$ directly related to 
the distance of closest approach $d_{{\rm min}}$ as 
$e_2 = M/2d_{{\rm min}} -1/12$. 
Then we obtain an addition formula for 3 points on the orbit in the form 
\begin{equation}
\frac{M}{2r(\phi_1 +\phi_2)} =\frac{1}{12} +
 F\left(F\left(\frac{M}{2r(\phi_1)}-\frac{1}{12},\frac{M}{2r(\phi_2)}-\frac{1}{12}\right),e_2\right)
\end{equation}
The same procedure for trapped orbits gives the same formula only with 
$e_1$ instead of $e_2$ where $e_1$ is related to the maximal 
attained distance $d_{{\rm max}}$ by $e_1 = M/2d_{{\rm max}} -1/12$.\\\\
In the absorbed case, the lack of additive constant in the 
argument of the Weierstrass function means that we can apply 
the addition formula directly to obtain algebraically simpler result
\begin{equation}
\frac{M}{2r(\phi_1 +\phi_2)} = \frac{1}{12} + F\left(\frac{M}{2r(\phi_1)} - \frac{1}{12},\frac{M}{2r(\phi_2)} - \frac{1}{12}\right)
\,. \end{equation}
In this case we can use the euclidean angle 
between the direction of the ray and $\phi$-direction $\psi$ which satisfies
\begin{equation}
\tan\psi = \frac{1}{r}\frac{dr}{d\phi}
\end{equation}
Then the addition formula can be written as 
\begin{equation}
\left(\frac{1}{r(\phi_1)}-\frac{1}{r(\phi_2)}\right)^2\left(2M(\frac{1}{r(\phi_1)}+\frac{1}{r(\phi_2)}+\frac{1}{r(\phi_1+\phi_2)})-1\right) = \left(\frac{\tan\psi(\phi_1)}{r(\phi_1)}-\frac{\tan\psi(\phi_2)}{r(\phi_2)}\right)^2
\,.\end{equation}
However, in this case this is not very useful 
since it is practically impossible to identify the 
line $\phi=0$ for such a choice. We could of course make a 
different choice, like $r(\phi=0)=R$ for some chosen $R$, 
but then the obtained addition formula would not be analytic 
anymore because we would need to find the corresponding additive 
constant $\phi_0$ given by the integral
\begin{equation}
\phi_0 = \int_{\frac{M}{2R}-\frac{1}{12}}^{\infty}\frac{dt}{\sqrt{4t^3 -t/12 -g_3}} 
\,.\end{equation}
\subsection{The deflection angle}
We  start from the equation for $u=2M/r$ which is
\begin{equation}
\left(\frac{du}{d\phi}\right)^2 = u^3 - u^2 +4M^2P =u^3 - u^2 -\mu^3+\mu^2
\,,\end{equation}
where $\mu=2M/r_0$, $r_0$ is the distance of closest approach, 
for scattering orbits. Then the deflection angle $\delta\phi$ is given by
\begin{equation}
\delta\phi = 2\int_0^{\mu}\frac{du}{\sqrt{u^3-u^2-\mu^3+\mu^2}} - \pi = 2I-\pi
\,.\end{equation}
The integral can be rewritten in terms of $x = u/\mu$ which gives
\begin{equation}
I(\mu) = \int_0^1\frac{dx}{\sqrt{(1-x^2)-(1-x^3)\mu}}
\,.\end{equation}
This integral can be expanded in the powers of 
$\mu$, for $\mu$ sufficiently small, as
\begin{equation}
I = \sum_{n=0}^{\infty}\frac{1}{4^n}{2n\choose n}\left(\int_0^1\left(\frac{1-x^3}{1-x^2}\right)^n\frac{1}{\sqrt{1-x^2}}dx\right)\mu^n \,.
\end{equation}
We are  only interested in small values of $\mu$ and this 
expansion clearly converges at least for $\mu<2/3$ 
since $(1-x^3)/(1-x^2)<3/2$ for $x\in (0,1)$. 
Calculating the  first three terms in this expansion  
 results in an  expansion for the deflection angle is (substituting for $\mu$)
\begin{equation}
\delta\phi = \frac{4M}{r_0}+3\left(\frac{5\pi}{4}-\frac{4}{3}\right)\frac{M^2}{r_0^2}+O(r_0^{-3})\,.
\end{equation}
Now, we have $P = 1/b^2$ and so define $\nu=2M/b$. Then
\begin{equation}
\nu^2 = \mu^2-\mu^3 \,,
\end{equation}
and working to the second order gives
\begin{equation}
\mu = \nu + \frac{1}{2}\mu^2 \,.
\end{equation}
Substituting into the expansion for $\delta\phi$ then gives
\begin{equation}
\delta\phi = 2\nu+\frac{15\pi}{16}\nu^2+O(\nu^3) = 
\frac{4M}{b} + \frac{15\pi}{4}\frac{M^2}{b^2}+O(b^{-3}) \,.
\end{equation}
which is the expansion of the deflection angle to second order in $1/b$.
\subsection{Angular sum in light triangles}

Because the Gauss curvature of the optical metric
restricted to the equatorial plane is negative, the
angular sum of a triangle made up of geodesics
must  less than $\pi$ unless the triangle encloses  the horizon
\cite{GW}.
One might hope to get a more precise statement using the addition formulae.  
To this end,  let $\Theta$ be the physical angle between the direction of 
the light and the $\phi$ - direction. Then
\begin{equation}
\tan\Theta = \frac{1}{\sqrt{1-\frac{2M}{r}}}\frac{1}{r}\frac{dr}{d\phi}=\sqrt{\frac{Pr^3 -r+2M}{r-2M}} \,.
\end{equation}
Note that this formula is valid for the Schwarzschild solution but
not the Kottler solution with non-vanishing cosmological constant
\cite{GWW}. 

\begin{figure}[htbp]
  \centering
    \includegraphics[scale=0.6]{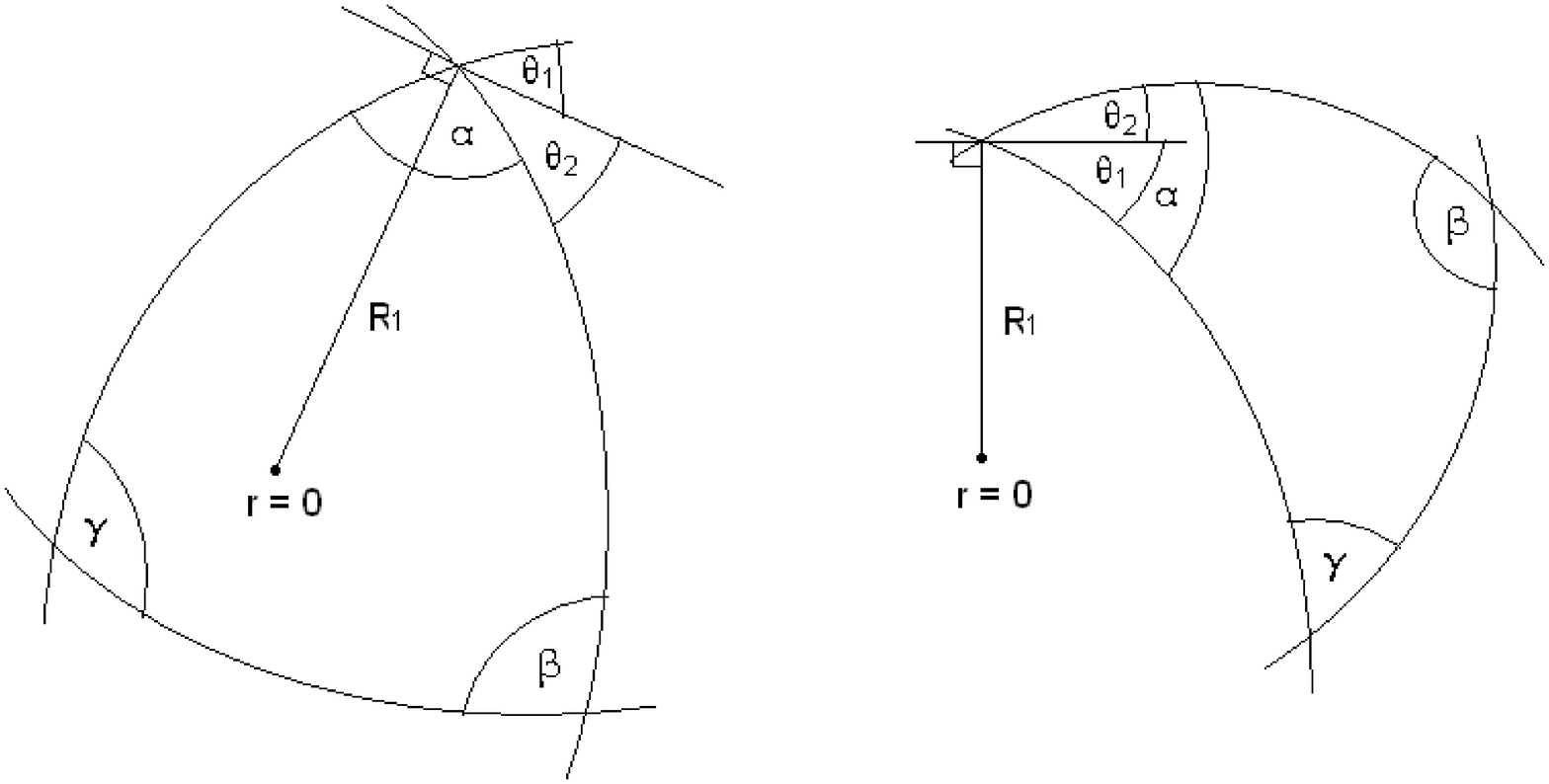}
  \label{Figure}
  \vspace{3mm}
  \centerline{\textbf{Figure 3} \emph{Light triangles}}
\end{figure}

Now consider 3 light rays, forming a triangle around the origin with $P_1$, $P_2$ and $P_3$ and vertices at the radial coordinate $R_1$, $R_2$ and $R_3$. To simplify the notation, define
\begin{equation}
W_{ij} = \sqrt{\frac{P_iR_j^3 -R_j+2M}{R_j-2M}}
\,.\end{equation}
Then from the Figure 3 it is clear that
\begin{eqnarray}
&& \alpha = \pi - \tan^{-1}W_{11} - \tan^{-1}W_{21}\,,\\ 
&& \beta = \pi - \tan^{-1}W_{22} - \tan^{-1}W_{32}\,,\\
&& \gamma = \pi - \tan^{-1}W_{13} - \tan^{-1}W_{33}\,.
\end{eqnarray}
Alternatively, if the origin is not inside of the triangle, then from the Figure 3 it follows that
\begin{eqnarray}
&& \alpha = \tan^{-1}W_{11} + \tan^{-1}W_{21}\,,\\
&& \beta = \pi - \tan^{-1}W_{22} - \tan^{-1}W_{32}\,,\\
&& \gamma = \tan^{-1}W_{13} + \tan^{-1}W_{33} \,.
\end{eqnarray}
Further analytical work in this general case doesn't seem to lead anywhere, 
because the distances $R_1$, $R_2$ and $R_3$ are not independent, 
but finding a formula for the relation between them is impossible. 
We can however consider a symmetric case with all $R$'s and $P$'s equal. 
Then its angles are given by
\begin{equation}
\alpha = \pi -2\tan^{-1}\sqrt{\frac{PR^3 -R+2M}{R-2M}} \,.
\end{equation}
\subsection{Gauss-Bonnet theorem}
An alternative approach to finding the angular deflection is 
using the Gauss-Bonnet theorem \cite{GW}. 
Consider the setup in the Figure 4. Then by the Gauss-Bonnet  theorem we have
\begin{equation}
\alpha + \pi + \int_AKdA = 2\pi \,.
\end{equation}
One of the way  to calculate this is 
transform the optical metric into the form
\begin{equation}
ds^2 = d\rho ^2+C(\rho)^2d\phi ^2 \,.
\end{equation}
Then 
\begin{equation}
KdA = -\frac{d^2C}{d\rho^2}d\rho d\phi \,,
\end{equation}
and so 
\begin{equation}
\int_AKdA = \int_{-\alpha/2}^{\alpha/2}\left[-\frac{dC}{d\rho}\bigg\vert_{r=\infty} + \frac{dC}{d\rho}\bigg\vert_{r=r(\phi)}\right]d\phi \,.
\end{equation}
Now, 
\begin{equation}
\frac{dC}{d\rho} = \frac{dC}{dr}\frac{dr}{d\rho} = \frac{r-M}{\sqrt{r}\sqrt{r-2M}} \,.
\end{equation}
Therefore we get
\begin{equation}
\int_{-\alpha/2}^{\alpha/2}\frac{r(\phi)-M}{\sqrt{r(\phi)}\sqrt{r(\phi)-2M}}d\phi = \pi \,,
\end{equation}
which holds for any scattering path. It doesn't seem to be very useful when it comes to evaluating $\alpha$ but it is an interesting expression. Rewriting this in terms of $r$ gives another interesting identity
\begin{equation}
\int_{r_0}^{\infty}\frac{r-M}{\sqrt{r}\sqrt{r-2M}\sqrt{Pr^4 - r^2 +2Mr}}dr = \frac{\pi}{2} \,.
\end{equation}
where $r_0$ is the distance of closest approach.

\begin{figure}[htbp]
  \centering
    \includegraphics[scale=0.6]{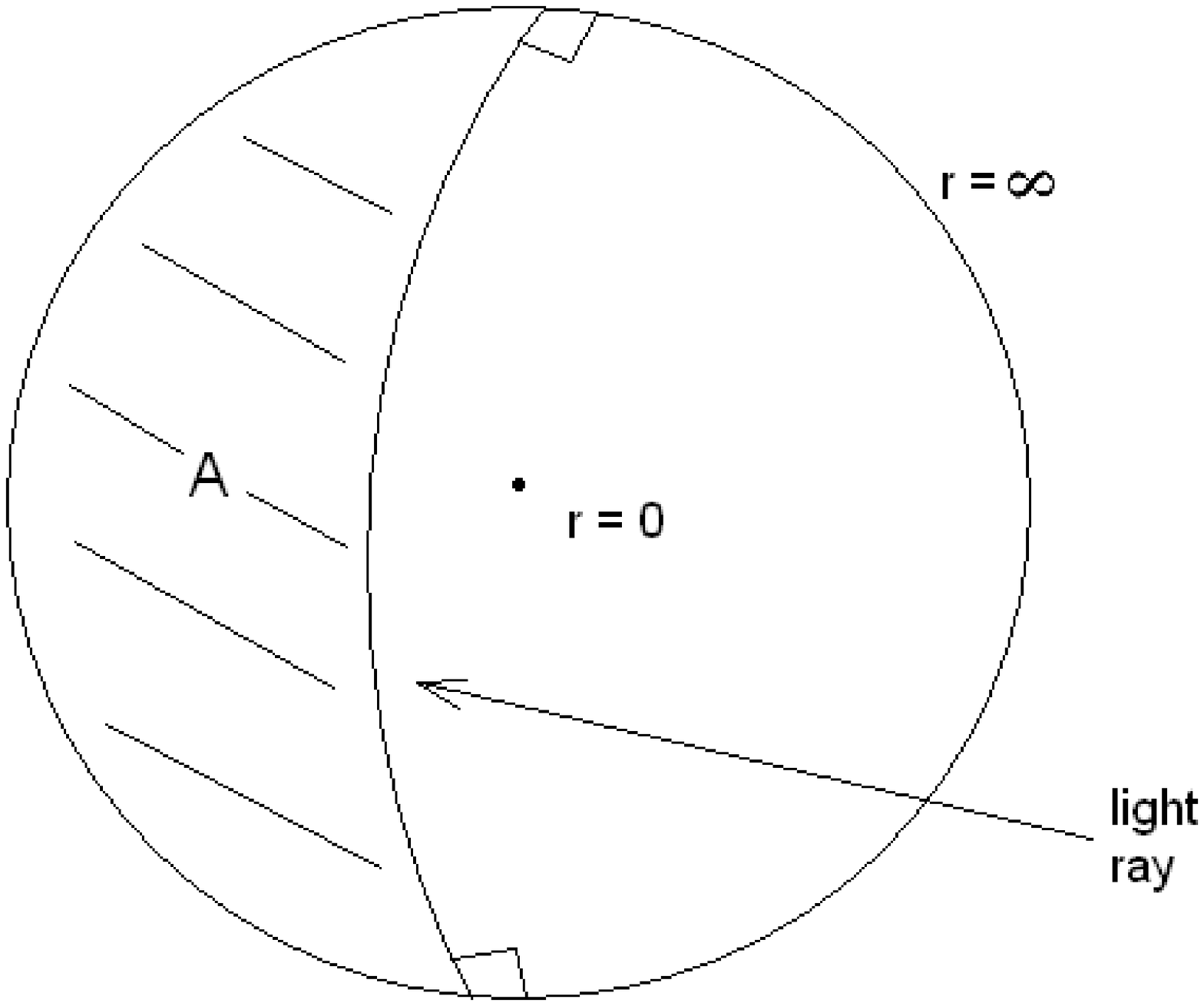}
  \label{Figure}
  \vspace{3mm}
  \centerline{\textbf{Figure 4} \emph{Scattering light ray}}
\end{figure}

\section{Further applications  of Weierstrass functions}
\subsection{Reissner Nordstr\o m null geodesics}
As mentioned in the introduction, these have been studied previously
using Weierstrass
functions  in \cite{Dabrowski}.
In this case, the relevant equation for $u = 1/r$ is 
\begin{equation}
\left(\frac{du}{d\phi}\right)^2 = P -u^2 +2Mu^3 -Q^2u^4 
\,. \label{this} \end{equation}
One may verify using
the formulae in \cite{GWW} or directly,
that  just as in the case of the Schwarzschild-de-Sitter
metrics, so with the Reissner Nordstr\o m  metrics 
that the cosmological constant does occur in (\ref{this}).
Thus some  of the results in \cite{Cruz}, which appeared
on the archive subsequently to the first version
of this paper,  follow directly from
the work of the present section.

The r.h.s.  of (\ref{this})  always has a 
real root so let $x_0$ to be one. Then define $s = u-x_0$. This gives
\begin{equation}
(s')^2 = As +Bs^2 +Cs^3 +Ds^4 \,, 
\end{equation}
where
\begin{eqnarray}
&&A = 6x_0^2M-2x_0-4x_0^3Q^2\,,\\&&
B = 6x_0M -1 -6x_0^2Q^2\,,\\&&
C = 2M -4x_0Q^2\,,\\&&
D = -Q^2 \,.
\end{eqnarray}
Now substitution $\psi = 1/s$ takes it into the form
\begin{equation}
(\psi ')^2 = A\psi^3+B\psi^2+C\psi+D \,, 
\end{equation}
and finally setting $\psi = 4y/A -B/3A$ gives
\begin{equation}
(y')^2 =4y^3 -g_2y-g_3  \,,
\end{equation}
where
\begin{eqnarray}
&& g_2 = \frac{B^2}{12} -\frac{AC}{4}\,,\\
&& g_3 = \frac{ABC}{48} -\frac{A^2D}{16} -\frac{B^3}{216} \,.
\end{eqnarray}
Therefore this time the solution will be given by 
\begin{equation}
\frac{1}{r(\phi)} = x_0 + \frac{3A}{12\wp(\phi + \xi_0) -B} \,,
\end{equation}
where $\xi_0$ is a complex constant. However, the more complicated relation 
between $r$ and $\wp$ and also many different constants make it 
algebraically very challenging to analyze the situation any further and 
find a suitable $\xi_0$ or addition formula similar to the Schwarzschild case. For that purpose, consider the equation for $r(\lambda)$ where 
$\lambda$ is an affine parameter of the path. This equation is
\begin{equation}
\frac{1}{L^2}\left(\frac{dr}{d\lambda}\right)^2 = P - \left(\frac{1}{r^2}-\frac{2M}{r^3}+\frac{Q^2}{r^4}\right)\equiv f(r) \,. 
\end{equation}
Now the motion is only possible in regions where $f(r)>0$. 
If these split into two disconnected ones, then that must be the case in which the r.h.s.  of the corresponding Weierstrass equation has 
3 real roots and we have scattering and trapped paths. If there is only one such region, we know we have the case where the above mentioned r.h.s. 
 has only 1 real root and we have absorbed paths. \\
First, the roots of $f'(r)$ are
\begin{equation}
r_{\pm} = \frac{3M\pm\sqrt{9M^2 -8Q^2}}{2} \,. 
\end{equation}
Physically we want $M^2 > Q^2$ and so $r_{\pm}\in\mathbb{R}^+$. 
Clearly the regions where $f(r)>0$ will be disconnected (and there will be 2) if $P>0$ and $f(r_+)<0$. Suppose that this is the case and consider scattering paths. let $a$ be the distance of closest approach. Then $a = e_i$ and we can write $y(\phi) = \wp(\phi +\omega_i)$ for some $i\in \lbrace1,2,3\rbrace$ because the path is symmetric. As before, this corresponds to choosing the line $\phi=0$ to be the axis of symmetry. We can compute $a$ as the largest root of $f(r)=0$ and so we obtain an addition formula
\begin{equation}
\frac{1}{12}\left(B +\frac{3A}{\frac{1}{r(\phi_1 +\phi_2)}-x_0}\right) = F\left(F\left(\frac{1}{12}\left(B +\frac{3A}{\frac{1}{r(\phi_1)}-x_0}\right),\frac{1}{12}\left(B +\frac{3A}{\frac{1}{r(\phi_2)}-x_0}\right)\right),a\right) \nonumber
\,.\end{equation}
For trapped orbits the same addition formula applies, only in that case $a$ is the largest attained distance and is given by the second largest root of $f(r)=0$.\\
If $f(r_+)>0$ then we have orbits that go in, miss the singularity, and continue to another asymptotically flat region of spacetime. These satisfy the same addition formula like the scattering ones.
In the case $P=0$ the equation can be integrated and the solution is
\begin{equation}
\frac{r}{\sqrt{2Mr-Q^2-r^2}} = \arctan(\phi -\phi_0)
\,,\end{equation}
where $\phi_0$ is the constant of integration. Another special solutions solutions can be found when
\begin{equation}
P = \frac{r_+^2-Q^2}{3r_+^4} \,,
\end{equation}
which is equivalent to $f(r_+)=0$ and corresponds to the situation when the two periods of the corresponding Weierstrass function become 
linearly dependent. This leads to a pair of solutions
\begin{equation}
r(\phi) = \frac{4ce^{\sqrt{c}\phi}}{-2be^{\sqrt{c}\phi}\pm(1+(b^2-4ac)e^{2\sqrt{c}\phi})} -r_+ \,,
\end{equation}
where
\begin{eqnarray}
&& a = \frac{r_+^2-Q^2}{3r_+^4}\,,\\
&& b = 4\frac{r_+^2-Q^2}{3r_+^3}\,,\\
&& c = 2\frac{r_+^2-Q^2}{r_+^2} -1 \,.
\end{eqnarray}
\subsection{5-D Schwarzschild null geodesics}
Here by 5-D it is meant 4 spatial dimensions. The relevant equation for $u=1/r$ in this case is
\begin{equation}
(u')^2 = 2Mu^4 -u^2 +P  \,,
\end{equation}
where $M$ is proportional to the five-dimensional mass.
There is an interesting self-duality here, which is in fact a special case of 
Bohlin-Arnold duality, in that when we write the equation in terms of $r$ we get
\begin{equation}
(r')^2 = Pr^4 -r^2 + 2M \,.
\end{equation}
which is exactly the same with the constants interchanged. First consider the equation for $u$. Substitution of 
 $u^2 = \frac{1}{2M}(y+\frac{1}{3})$ will take it into a form 
\begin{equation}
(y')^2 = 4y^3 -g_2y - g_3 \label{tthis} 
\,,\end{equation} 
where
\begin{eqnarray}
&& g_2 = \frac{4}{3} -8MP \,, \\
&& g_3 = \frac{8}{3}\left(\frac{1}{9}-MP\right) \,.
\end{eqnarray} As usual, the r.h.s.  of (\ref{tthis})  has 3 real
roots if $g_2>0$ and $g_3^2<(g_2/3)^3$. The first condition is
$MP<1/6$ while the second is $8(MP)^3 -(MP)^2<0$. Therefore we have 4
real roots if $MP<1/8$. Also, note that the r.h.s.of  equation (32)
always has root $-1/3$ and expanding it into a  power series around
this point quickly shows that in fact $e_3 = -1/3$.  Finally, the point
$y=-1/3$ corresponds to $r=\infty$. Hence in this case, with
$\omega_1,\omega_3$ defined as before, we have 2 classes of solutions,
depending on the initial conditions, scattering or trapped. The case
of trapped paths is exactly the same as before, with the same addition
formula for $y$ and $y(\phi) = \wp(\phi +\omega_1)$. However, the
scattering case is more interesting in 5D. This is because now the point
$\omega_3$ in the $\mathbb{C}$-plane corresponds to $r=\infty$ and so
we can write the solution as $y(\phi) = \wp(\phi +\omega_3)$ where the
line $\phi=0$ is in the direction of the ray incoming from $\infty$
and $\phi\in[0,2\omega_1]$.\\ Things are even simpler when we solve
the equation for $r$ directly. The substitution $r^2 =
\frac{1}{P}(y+\frac{1}{3})$ takes it into the equation (32) but now
the difference is that point $0$ corresponds to $r=\infty$ and so we
can write the (scattering) solution simply as 
\begin{equation}
r(\phi) = \frac{1}{\sqrt{P}}\sqrt{\wp(\phi)+\frac{1}{3}} \,,
\end{equation}
and the addition formula in this case is simply
\begin{equation}
r(\phi_1+\phi_2) = \frac{1}{\sqrt{P}}\sqrt{F\left(P(r(\phi_1))^2-\frac{1}{3},P(r(\phi_2))^2-\frac{1}{3}\right)+\frac{1}{3}} \,.
\end{equation}
Finally, if $MP>1/8$ then we have only 1 root of the r.h.s. of the Weierstrass equation and thus absorbing paths for which the solution is 
\begin{equation}
\frac{1}{(r(\phi))^2} = \frac{1}{2}\left(\wp(\phi+\omega_1)+\frac{1}{3}\right)
\,,\end{equation}
where again the line $\phi=0$ is given by the direction of the ray incoming from $\infty$.\\ As before, we can get several special solutions by imposing $g_2^3 = 27g_3^2$ which in this case gives $MP=0$ or $MP=1/8$. In the case $MP=0$ we get a special {\it circular}  solution
\begin{equation}
r(\phi) = \sqrt{2M}\cos\phi  \,,
\end{equation}
while in the case $MP=1/8$ we get
\begin{equation}
r(\phi) = 2\sqrt{M}(\tanh(\phi/\sqrt{2}))^{\pm 1} \,.
\end{equation}
\subsection{Duality and 7-D Schwarzschild null geodesics}
By  Bohlin-Arnold duality \cite{Arnold}, if we have a particle moving in Newtonian potential $V \propto r^{2p-2}$ and following trajectory $r(\phi) = f(\phi)$ then there will be a particle with accordingly modified energy moving in a potential $\widehat{V}\propto r^{\frac{2-2p}{p}}$ following trajectory $r(\phi) = f(\phi)^p$. In this case, if $V = -kr^{2p-2}$ and particle has energy $E$ then $\widehat{V} = -Er^{\frac{2-2p}{p}}$ and $\widehat{E}=k$. \\
Null geodesics in the $(n+1)$D Schwarzschild geometry correspond to 
Newtonian motion in a $r^{-n}$ potential and so the duality applies to these geodesics as well. As we already saw, the 5-D corresponds 
to the case $p=-1$ and is self dual. A quick check reveals 
that the case $p=-1/2$ gives a duality between null geodesics in 7-D and 4-D. \\
Given the potential $V = -kr^{-n}$,   Newton's equation of motion is
\begin{equation}
(r')^2 = \frac{2E}{L^2}r^4 -r^2 +\frac{2k}{L^2}r^{4-n} \,.
\end{equation}
The equation for null geodesics in 4-D is
\begin{equation}
(r')^2 = Pr^4 -r^2 +2Mr \,,
\end{equation}
and in 7-D it is
\begin{equation}
(r')^2 = Pr^4 -r^2 +2Mr^{-2} \,.
\end{equation}

So, under the duality with $p=-1/2$ we have $E\leftrightarrow k$ and thus 
$P\leftrightarrow 2M $. Therefore if we have 4-D Black Hole  with mass $M$ and 
light with $(E/L)^2=P$ following the  trajectory $r(\phi)$ and 7-D Black Hole 
with mass $P/2G$ and light with $(E/L)^2 = 2M$ 
following the  trajectory $r = f(\phi)$ then
\begin{equation}
r(\phi) = \left(\frac{1}{f(\phi)}\right)^2 \,.
\end{equation}
Making the  substitution $r^2 = y/P +1/(3P)$ in the equation for $r$ in 
the 7-D case takes it into the Weierstrass form 
with $g_2 = 4/(3MP^2)$ and\\ $g_3 = 8/27 -8M_6P^2$. 
Therefore the orbits in 7-D satisfy 
\begin{equation}
r(\phi) = \frac{1}{\sqrt{P}}\sqrt{\wp(\phi +C)+\frac{1}{3}} \,.
\end{equation}
In this case, the formula for scattering paths looks especially simple, it is
\begin{equation}
r(\phi) = \frac{L}{E}\sqrt{\wp(\phi)+\frac{1}{3}} \,.
\end{equation}
By Bohlin-Arnold duality, the special solutions in 7-D corresponding to the special solutions in 4-D given by $P=1/27$ have
\begin{equation}
P = \frac{2}{\sqrt{54M}}
\end{equation}
where $M$ is proportional to the mass of 7-D black hole. The corresponding special solutions thus are
\begin{equation}
r(\phi) = \sqrt[4]{54M}\sqrt{\frac{1}{3}-\frac{1}{1\pm\cosh(\phi)}} \,.
\end{equation}
\subsection{Ellis Wormhole null geodesics}
\subsubsection{Qualitative description}
The  Ellis wormhole, is an ultra static  solution of the Einstein equations
coupled to a massless scalar field. While not necessarily 
physically very  realistic,
has been used in studies of gravitational lensing \cite{DeySen}. It  
 has  the metric 
\begin{equation}
ds^2 = -dt^2+dr^2+r(r-2M)(d\theta ^2+sin^2\theta d\phi^2)
\end{equation}
Because $g_{00}=-1$, the physical spatial metric and the optical spatial
metric coincide. 
Setting $t=0$, $\theta = \frac{\pi}{2}$ gives the optical metric 
on the equatorial plane. 

If we set $\sqrt{x^2+y^2} = \sqrt{ (r-M)^2 -M^2}$   
we may isometrically embed into ${\Bbb E} ^3$ with coordinates
$(x,y,z)$ as the  surface of revolution
\begin{equation}
\sqrt{x^2+y^2}  = M \cosh \frac{z}{M} \,, \qquad r= M(1+  \sinh \frac{z}{M} ) 
  \label{catenoid} .
\end{equation}
Note that (\ref{catenoid}) is a {\it catenoid}. 
This may be compared with the well known {\it Flamm paraboloid} 
which gives an isometric embedding
of the physical equatorial plane geometry of the 
Schwarzschild metric 
\begin{equation}
\sqrt{x^2+y^2} = 2M + \frac{z^2}{8M} \,,\qquad r=\sqrt{x^2+y^2}  \,.
\end{equation}
It is also possible to isometrically embed the Schwarzschild optical metric 
(\ref{opt}) into Euclidean space but the formulae are more complicated:
\begin{equation}
\sqrt{x^2+y^2}  = \frac{r}{\sqrt{1-\frac{2M}{r}}} 
 \,, \qquad z= \int^r  \sqrt{\frac{M}{r}(4-9\frac{M}{r})}
 \bigl (1-\frac{2M}{r} \bigr )^{-\frac{3}{2}} \,.  
\end{equation}

If we let $u=\frac{1}{r-M}$ then the equation of null geodesic is
\begin{equation}
(u')^2 = (\xi -1)M^2u^4 +(2\xi -1)u^2 +\frac{\xi}{M^2} \label{u eq}
\,,\end{equation}
where $\xi = M^2E^2/L^2$. Note that this equation does not distinguish between $r$ and $2M-r$ for $r\in[0,M]$. Before turning to 
the Weierstrass functions, we give a  qualitative analysis 
of the null geodesics. Going back to the equation for $r$ gives
\begin{equation}
(r')^2 = \frac{\xi}{M^2}r^4 -\frac{4\xi}{M}r^3+(8\xi-1)r^2+2M(1-4\xi)r+2M^2(2\xi-1)\equiv f(r) \,.
\end{equation}
The roots of $f(r)$ have a very simple form, they are
\begin{eqnarray}
&& r = (1\pm i)M\,,\\
&& r = M\left(1\pm\sqrt{\frac{1}{\xi}-1}\right) \,.
\end{eqnarray}
Extremal points of $f(r)$, roots of $f'(r)$ also have a simple form, they are
\begin{eqnarray}
&& r = M\,,\\
&& r = M\left(1\pm\sqrt{\frac{1}{2\xi}-1}\right) \,.
\end{eqnarray}
From these result it follows that if
\begin{itemize}
\item $\xi\in(0,1/2)$ then $f(r)$ has 1 positive real root $M(1+\sqrt{1/\xi -1})$ and 3 local extrema, all with value smaller than this root.
\item $\xi\in(1/2,1)$ then $f(r)$ has 2 positive real roots $M(1\pm\sqrt{1/\xi -1})$ and 1 global extremum (minimum) at $r=M$.
\item $\xi\in(1,\infty)$ then $f(r)$ has no real roots and 1 global extremum (minimum) at $r=M$.
\end{itemize}
This shows that if
\begin{itemize}
\item $\xi\in(0,1/2)$ There are \textbf{only scattering orbits} with the distance of closest approach $M(1+\sqrt{1/\xi -1})$
\item $\xi\in(1/2,1)$ There are both \textbf{scattering and trapped orbits} with the distance of closest approach $M(1+\sqrt{1/\xi -1})$ and the largest attained distance $M(1-\sqrt{1/\xi -1})$, respectively.
\item $\xi\in(1,\infty)$ There are \textbf{only absorbing orbits} that is orbits incoming from $\infty$ that hit $r=0$.
\end{itemize}
There is an important point here. Suppose that we  wanted to express 
$r$ in terms of some Weierstrass function. The only way how 
to convert the full quartic into cubic is to substitute 
$r = x+r_0$ with $r_0$ being a root of $f(r)=0$ and then $s = 1/x$. 
If this approach is to be useful, we want $r_0\in\mathbb{R}$,
 since otherwise,  we would be looking for complex solution of the 
Weierstrass equation and the imaginary part $C$ in $\wp(\phi+C)$ 
would not be half-period anymore but rather some analytically 
incalculable number and so this approach would not be useful at all. 
But $f(r)$ \textbf{has no} real root in the case of absorbing paths and this 
foretells problems when treating this case.
\subsubsection{Weierstrass function approach}
First we  make the substitution $u^2 = 1/x$ in the equation $(\ref{u eq})$ which takes it into the form
\begin{equation}
\frac{1}{4}(x')^2 = (\xi -1)x +(2\xi-1)x^2 +\frac{\xi}{M^2}x^3 \,.
\end{equation}
Then the substitution 
\begin{equation}
x = \frac{M^2y}{\xi}+\frac{M^2(1-2\xi)}{3\xi} 
\end{equation}
takes it into Weierstrass form
\begin{equation}
(y')^2 = 4y^3 -g_2y-g_3 \label{weier} \,,
\end{equation}
where
\begin{eqnarray}
&& g_2 = \frac{4}{3}(1-\xi+\xi^2)\,,\\
&& g_3 = \frac{4}{27}(2-3\xi-3\xi^2+2\xi^3) \,.
\end{eqnarray}
Note that $g_2>0$ $\forall\xi$ and that 
\begin{equation}
\left(\frac{g_2}{3}\right)^3-g_3^2 = \frac{16}{27}(\xi-1)^2\xi^2>0
\,,\end{equation}
unless $\xi=0,1$. Setting $\xi=0$ in eq. ($\ref{u eq}$) shows that 
this case is not possible. The case $\xi=1$ gives 2 analytical solutions
\begin{equation}
r_{\pm}(\phi) = M\left(1\pm\frac{1}{\sinh{\phi}}\right)
\,,\end{equation}
where $r_+$ comes from $\infty$, $r_-$ comes from $r=0$ and both are 
approaching $r=M$, but never reaching it.
For other values of $\xi$ the r.h.s. 
of equation ($\ref{weier}$) has 3 real roots $e_1>e_2>e_3$ where
\begin{eqnarray}
&& e_1 = \max\left(\frac{2-\xi}{3},\frac{2\xi-1}{3}\right)\,,\\
&& e_2 = \min\left(\frac{2-\xi}{3},\frac{2\xi-1}{3}\right)\,,\\
&& e_3 = -\frac{1}{3}(1+\xi) \,.
\end{eqnarray}
Now we  will analyze the separate cases. Suppose that:
\begin{itemize}
\item \underline{$\xi\in(0,1/2)$}. Then
\begin{eqnarray}
&& e_1 = \frac{2-\xi}{3}\,,\\
&& e_2 = \frac{2\xi-1}{3}\,,\\
&& e_3 = -\frac{1}{3}(1+\xi) \,.
\end{eqnarray}
As a consistency check, one can verify that plugging 
$y=e_1$ into the expression $r=r(y)$ indeed gives
 $r=M(1+\sqrt{1/\xi -1})$ as it should. Also, 
the point $r=\infty$ corresponds to the  point $y=\infty$ 
and so the solution for the scattering orbits in this case is
\begin{equation}
\frac{r(\phi)}{M} = 1+\frac{1}{\sqrt{\xi}}\sqrt{\wp(\phi)+\frac{1-2\xi}{3}}
\,,\end{equation}
where the line $\phi=0$ is in the direction of the ray incoming from 
$\infty$ and $\phi\in(0,2\omega_1)$. 
Note that this solution always stays above $r=2M$.\\
The point $r=0$ corresponds to $y=(5\xi-1)/3$ which is in 
this case in an unphysical region and so in accordance with Section 
1.1 we only have scattering solutions in this case.
\item \underline{$\xi\in(1/2,1)$}. Then
\begin{eqnarray}
&& e_1 = \frac{2-\xi}{3}\,,\\
&& e_2 = \frac{2\xi-1}{3}\,,\\
&& e_3 = -\frac{1}{3}(1+\xi) \,.
\end{eqnarray}
But now the scattering solutions penetrate into the region 
$M<r<2M$ and so I have to be careful here because $r(y)$ is multivalued
\begin{equation}
r = M\left(1\pm \frac{1}{\sqrt{\xi}}\sqrt{y+\frac{1-2\xi}{3}}\right)
\,.\end{equation}
This only becomes a problem once the orbit crosses $r=2M$ and so 
we didn't   have to worry about it in the previous case $\xi<1/2$.\\
In this case $(5\xi -1)/3>e_1$ and $r(5\xi -1)/3) = 0$ or $2M$. 
For orbits incoming from $\infty$ we clearly have to choose 
$r(5\xi -1)/3) = 2M$ because $r=M$ is inaccessible. 
\\Also $r(e_1) = M(1\pm\sqrt{1/\xi -1})$ and for the same reason 
we have to choose + for orbits incoming from $\infty$. Hence as before
\begin{equation}
\frac{r(\phi)}{M} = 1+\frac{1}{\sqrt{\xi}}\sqrt{\wp(\phi)+\frac{1-2\xi}{3}}
\,,\end{equation}
where again the line $\phi=0$ is in the direction of the ray incoming from $\infty$ and $\phi\in(0,2\omega_1)$.\\
What is left are orbits trapped in the region $r<M(1-\sqrt{1/\xi -1})$. For these we need to choose minus signs in the above equations and so we get
\begin{equation}
\frac{r(\phi)}{M} = 1-\frac{1}{\sqrt{\xi}}\sqrt{\wp(\phi+\omega_1)+\frac{1-2\xi}{3}} \,,
\end{equation} 
where now the additive constant in the argument of the Weierstrass function is necessary. This choice corresponds to setting the line $\phi=0$ to be the axis of symmetry and $\phi\in(-\beta,\beta)$ where
\begin{equation}
\beta = \omega_1 - \int_{(5\xi-1)/3}^{\infty}\frac{dt}{\sqrt{4t^3-g_2t-g_3}} \,.
\end{equation} 
\item  \underline{$\xi\in(1,\infty)$}. Then
\begin{eqnarray}
&& e_1 = \frac{2\xi-1}{3}\,,\\
&& e_2 = \frac{2-\xi}{3}\,,\\
&& e_3 = -\frac{1}{3}(1+\xi) \,.
\end{eqnarray} We know that in this case all orbits are incoming from
$\infty$ and reach $r=0$. Both $y=e_2$ and $y = e_3$  correspond to
unphysical (complex) $r$ and this time, $y = e_1$ corresponds to $r=M$
without any ambiguity. Suppose we have an orbit starting at
$\infty$. $(5\xi -1)/3>e_1$ and so we need to choose plus sign in the
relation $r(y)$.\\ Thus $r(y = \infty) =\infty$, then $r(y = (5\xi
-1)/3) = 2M$ and finally we reach $r(y=e_1)=M$. But if we continued
the same Weierstrass function solution now $r$ would begin to increase
again, which we know is unphysical. Therefore we need to switch the
branches and continue with minus sign in the relation $r(y)$ so that
we reach $r(y = (5\xi -1)/3) = 0$. Now there is no way of continuing the solution and we need to start a new one, first using a minus sign and
then a plus sign on its journey from $r=0$ to $r=\infty$. Therefore an
orbit going from $\infty$ to $r=0$ travels a total angle
$\omega_1+\beta$ and satisfies
\begin{eqnarray}
&& \frac{r(\phi)}{M} = 1+\frac{1}{\sqrt{\xi}}\sqrt{\wp(\phi)+\frac{1-2\xi}{3}}\hspace{1cm} {\rm for}\hspace{2mm} \phi\in(0,\omega_1)\,,\\
&& \frac{r(\phi)}{M} = 1-\frac{1}{\sqrt{\xi}}\sqrt{\wp(\phi)+\frac{1-2\xi}{3}}\hspace{1cm} {\rm for}\hspace{2mm} \phi\in(\omega_1,\beta)\,,
\end{eqnarray}
where again line $\phi=0$ is in the direction of the ray incoming from $\infty$.
\end{itemize}
\underline{General remarks}\\
(i) Note that the scattering solutions depend directly on $\wp(\phi)$ and so the addition formula for Weierstrass functions can be applied directly.\\
(ii) The same is true for the absorbing one, however we need to be careful to stay in the region $\phi\in(0,\omega_1)$ or $\phi\in(\omega_1,\beta)$ when applying it.
\subsubsection{Angle of deflection in the scattering case}
The equation for $u$ can be factorized as
\begin{equation}
(u')^2 = (1+M^2u^2)\left(\frac{\xi}{M^2}+(\xi-1)u^2\right) \,.
\end{equation}
Now, the distance of closest approach is $r_0 = M+M\sqrt{1/\xi -1}$, which corresponds to 
\begin{equation}
u_0 = \frac{1}{M}\sqrt{\frac{\xi}{1-\xi}} \,.
\end{equation}
Let $I$ be half of the angle $\phi$ travelled by the light. 
\begin{equation}
I = \int_0^{u_0}\frac{du}{\sqrt{1+M^2u^2}\sqrt{\xi/M^2+(\xi-1)u^2}} \,.
\end{equation} 
Making the  substitution $u = u_0t$, we have 
\begin{equation}
I = \frac{1}{M}\sqrt{\frac{\xi}{1-\xi}}\int_0^1\frac{dt}{\sqrt{1+t^2\frac{\xi}{1-\xi}}\sqrt{\frac{\xi}{M^2}+\frac{\xi}{M^2}t^2}}=\int_0^1\frac{dt}{\sqrt{1-t^2}\sqrt{1-(1-t^2)\xi}} \label{int} \,.
\end{equation}
Write $f(t,\xi)$ for the final integrand above. It is 
straightforward to differentiate $f$ n times w.r.t. $\xi$ and the result is
\begin{equation}
\frac{\partial^n f}{\partial\xi^n} = (1-t^2)^{\frac{2n-1}{2}}\frac{(2n-1)!!}{2^n}\frac{1}{(1-(1-t^2)\xi)^{\frac{2n+1}{2}}} \,.
\end{equation} 
Therefore we can expand $f$ as
\begin{equation}
f(t,\xi) = \sum_{n=0}^{\infty}(1-t^2)^{\frac{2n-1}{2}}\frac{(2n-1)!!}{n!}2^{-n}\xi ^n.
\end{equation}
Scattering orbits exist for $\xi\in(0,1)$ and for this range of values of $\xi$ the sum converges uniformly (for example by straightforward application of the Weierstrass M-test) and therefore we can write
\begin{equation}
I = \sum_{n=0}^{\infty}\left(\frac{(2n-1)!!}{n!}2^{-n}\xi 
^n\int_0^1(1-t^2)^{\frac{2n-1}{2}}\right) \,.
\end{equation}
The integral in this sum can be computed by hand, one  way is
as follows. The volume $V_n$  of an  $n$-dimensional ball 
\begin{equation}
V_n = \frac{\pi^{n/2}}{\Gamma\left(1+\frac{n}{2}\right)}
\end{equation}
Therefore
\begin{equation}
\int_0^1(1-t^2)^{\frac{2n-1}{2}} = \frac{V_{2n}}{2V_{2n-1}} = \frac{1}{2}\frac{\pi^n}{\Gamma(n+1)}\frac{\Gamma\left(n+\frac{1}{2}\right)}{\pi^{n-1/2}} = \frac{\sqrt{\pi}}{2n!}\Gamma\left(n+\frac{1}{2}\right)\,.
\end{equation}
Now using the identity
\begin{equation}
\Gamma\left(n+\frac{1}{2}\right) = (2n-1)!!2^{-n}\sqrt{\pi}
\end{equation}
gives
\begin{equation}
I = \frac{\pi}{2}\sum_{n=0}^{\infty}\left(\frac{(2n-1)!!}{n!}\right)^22^{-2n}\xi ^n
\,.\end{equation}
This can be further simplified using the identity 
$(2n-1)!!n! = (2n)!2^{-n}$ to give
\begin{equation}
I = \frac{\pi}{2}\sum_{n=0}^{\infty}{2n\choose n}^22^{-4n}\xi ^n
\,.\end{equation}
Now the angle of deflection 
$\delta\phi$ is given by $\delta\phi=\pi - 2I$ and so
\begin{equation}
\delta\phi = \pi-\pi\sum_{n=0}^{\infty}{2n\choose n}^22^{-4n}\xi ^n
\,.\end{equation}
The first few terms of this expansion are
\begin{equation}
\delta\phi = -\frac{\pi}{4}\xi -\frac{9\pi}{64}\xi^2-\frac{25\pi}{256}\xi^3-\frac{1225\pi}{16384}\xi^4-\frac{3969\pi}{65536}\xi^5 - \frac{53361\pi}{1048576}\xi^6 -\ldots
\end{equation}
with $\xi = (M/b)^2$.\\
We have  also tried expanding the deflection angle in terms of $\mu =
M/r_0$ 
following  \cite{DeySen}. Substituting
\begin{equation}
\xi = \frac{1}{1+\left(\frac{1}{\mu}-1\right)^2}
\end{equation}
into the integral ($\ref{int}$) and expanding in the powers of $\mu$, 
using Mathematica,  the first few terms  are
\begin{equation}
\frac{\delta\phi}{\pi} = -\frac{1}{4}\mu^2-\frac{1}{2}\mu^3-\frac{41}{64}\mu^4-\frac{9}{16}\mu^5-\frac{25}{256}\mu^6+\frac{37}{128}\mu^7+\frac{11959}{16384}\mu^8+\frac{1591}{2048}\mu^9+\frac{13311}{65536}\mu^{10}-\frac{29477}{32768}\mu^{11}-\ldots \,. 
\end{equation}
This expansion is not very useful , since the coefficients 
don't seem to be decreasing very fast, the coeff. of $\mu^{11}$ is almost 1.\\
Note that this expansion is completely different from the one given in  
\cite{DeySen}.

\section{Conclusion}
In this paper, we have used Weierstrass elliptic functions to give
a full description and  and classification of 
 null geodesics in Schwarzschild spacetime. We then  used this 
description to derive some  analytical formulae 
connecting three points on these geodesics and found second 
order expansion of the deflection angle in the scattering case. 
Finally, we  derived some properties of light triangles in this spacetime and used the Gauss-Bonnet theorem to derive a quantity which gives the same answer when integrated along a scattering geodesic, independently of the geodesic in question.\\
We  then showed that the Weierstrass elliptic function formalism can
also be used to describe other more exotic  spacetimes such as 
Reissner-Nordstr\o m null geodesics and Schwarzschild null geodesics in spacetimes with spatial dimensions 4 and 6. In all these cases, the elliptic function approach allows one to find the special case analytical solutions with ease (simply by looking at the values of parameters for which the elliptic function in question collapses into a periodic one).\\
Finally we  applied the formalism to describe the null geodesics of the Ellis wormhole and found an expansion for the angle of deflection in this case.

After the appearance of the first verson of this paper on the archive,
Betti Hartmann pointed out to us  that our results may easily extended to the 
case of a Schwarzschild black hole pierced by an infinitely cosmic string 
studied in \cite{Hackmann}. One need only replace the variable $\phi$   
by $\delta \phi$ , where $0 < \delta \leq 1 $ is the deficit parameter.
Similar remarks to the other metrics studied in this paper. 
 
\section{Acknowledgement} The work of M.~V.~ was supported by a 
 Trinity College Summer Studentship.


\end{document}